\documentstyle[aps,prb,psfig,twocolumn,floats]{revtex}

\newcommand{\jour}[4]{#1\ \textbf{#2},\ #3\ (#4)}
\newcommand{\book}[3]{\textit{#1} (#2,\ #3)}
\newcommand{\inbook}[4]{in \textit{#1}, edited by\ #2\ (#3,\ #4)}
\newcommand{\Rref}[1]{Ref.~\onlinecite{#1}}
\newcommand{\Eref}[1]{Eq.~(\ref{#1})}
\newcommand{\Fref}[1]{Fig.~\ref{#1}}
\newcommand{\Tref}[1]{Table~\ref{#1}}
\newcommand{\etal}{\textrm{et al.}}
\newcommand{\zin}{Z_{\text{In}}}
\newcommand{\inongaas}{In/GaAs(001)-$c(4\times4)$}
\newcommand{\gaas}{GaAs(001)-$c(4\times4)$}
\newcommand{\eps}{\varepsilon}
\newcommand{\kb}{k_{_{\text{B}}}}
\newcommand{\PRB}{Phys. Rev. B}
\newcommand{\PRL}{Phys. Rev. Lett.}
\newcommand{\APL}{Appl. Phys. Lett.}
\newcommand{\SSci}{Surf. Sci.}

\begin{document}

\wideabs{     
\title{Effect of strain on surface diffusion in semiconductor heteroepitaxy}

\author{Evgeni Penev,
         Peter Kratzer, and 
         Matthias Scheffler
}

\address{%
  Fritz-Haber-Institut der Max-Planck-Gesellschaft, 
  Faradayweg 4-6, 14195 Berlin-Dahlem, Germany
}

\date{Submitted 25 January 2001}
 
\maketitle
\begin{abstract} 
  We present a first-principles analysis of the strain
  renormalization of the cation diffusivity on the GaAs(001)
  surface. For the example of \inongaas\ it is shown that the binding
  of In is increased when the substrate lattice is expanded. The
  diffusion barrier $\Delta E(\eps)$ has a non-monotonic strain
  dependence with a maximum at compressive strain values ($\eps < 0$),
  while being a decreasing function for any tensile strain ($\eps >
  0$) studied.  We discuss the consequences of spatial variations of
  both the binding energy and the diffusion barrier of an adatom
  caused by the strain field around a heteroepitaxial island. For a
  simplified geometry, we evaluate the speed of growth of two
  coherently strained islands on the GaAs(001) surface and identify a
  growth regime where island sizes tend to equalize during growth due
  to the strain dependence of surface diffusion.  
\end{abstract}

\pacs{68.35.Fx, 68.65.+g, 71.15.Nc}
} 

\section{Introduction}
\label{intro}

In recent years, the heteroepitaxial growth of lattice-mismatched
semiconductor systems has attracted substantial interest. For a number
of systems (e.g. Ge/Si, InAs/GaAs), it could be shown that under
appropriate experimental conditions the deposited material forms
elastically strained, dislocation-free (coherent) three-dimensional (3D)
islands.~\cite{experimental,Leonard:94} In addition, these islands
often show a rather narrow size distribution, in particular in the
higher layers of a stacked 3D array of islands obtained from repeated
deposition of heteroepitaxial films separated by spacer layers.
This feature is essential for the usefulness of these nanostructures
as quantum dots (QDs) and for their envisaged application in future
optoelectronic devices.~\cite{Ledentsov98} Considerable theoretical
efforts have been made in order to rationalize the observed
regularities in island sizes and ordering.  Some approaches have
attempted to describe the islands as equilibrium
structures.~\cite{Priester95,Shchukin:99,Wang:99} As an alternative
explanation, the role of \textit{kinetics} for the growth of
heteroepitaxial islands has been
emphasized.~\cite{Dobbs:97a,Dobbs:97b,Koduvely:99} It is possible that
intrinsic features of the kinetics of the growth process give rise to
regular structures.  For instance, self-limiting effects in strained
island growth could result in a preferred island size, either due to a
limitation in material supply,\cite{Chen96} or due to nucleation
barriers in the growth of the islands' side facets.~\cite{Jesson98}
This perspective motivated an intense theoretical effort towards
better understanding the underlying microscopic processes in molecular
beam epitaxy (MBE), e.g. deposition, diffusion and nucleation.
First-principles calculations have already been applied to study
different aspects of cation diffusion on unstrained compound
semiconductor surfaces.~\cite{Kley,LePage}
Up to now, however, the impact of strain on the diffusion process
still remains elusive, although attempts to make its effect clearer
date back to the last decade.~\cite{Ghaisas}

We illustrate the importance of strain for typical heteroepitaxial
systems (e.g. Ge/Si, InAs/GaAs) for two situations: First, during
growth of free-standing heteroepitaxial islands, the islands
themselves are under compressive strain, whereas the substrate beneath
the island is expanded.  As a consequence of this expansion, the
substrate surface \textit{around} an island is under compressive
strain (see, e.g. \Rref{Moll98}).  Supply of further material to the
growing island is governed by diffusion through this ring-shaped area
of compressive strain.  A suppressive effect of strain on diffusion
could slow down the growth of larger islands.  Secondly, in the growth
of three-dimensional (3D) stacked arrays of QDs, the buried islands
act as stressors causing tensile strain on the capping layer in the
regions above them.~\cite{Xie:95,Schmidt:00} Again, this may affect
the growth kinetics of the next layer of islands to be formed on the
capping layer. Thus, in different stages of growth of a nanostructure,
different regimes of strain may come into focus.

From the limited published data about strain effects on diffusion, it
appears surprising that compressive surface strain could lead to a
self-limiting effect on the island growth.  First-principles
calculations for diffusion on close-packed metal surfaces, in
particular Ag/Ag(111), \Rref{Ratsch}, have demonstrated that
compressive strain increases the adatom diffusivity by reducing the
diffusion barrier.  Schroeder and Wolf~\cite{Schroeder:97} have
extended this finding to diffusion on (001) surfaces of
simple cubic, fcc, and bcc lattices.  Recent molecular dynamics (MD)
simulations using empirical potentials showed the same trend for Ga,
In, and As adatom diffusion on a $(2\times1)$-reconstructed 
GaAs(001) surface.~\cite{Matthai} These results also comply with an earlier
study of Ga kinetics on the strained GaAs(001) surface.~\cite{Ghaisas}
A different strain dependence of diffusion was found, however, for Si
adatom~\cite{Roland:92,Spjut:94} and dimer~\cite{Zoethout:00}
diffusion on the Si(001) surface, where tensile strain leads to an
overall decrease in the diffusion barriers.  Yet, the majorities of
the theoretical studies on semiconductor systems provide only scarce
quantitative information about the influence of strain on the
diffusion process.

The aim of this article is two-fold: First, we report the results of
density-functional theory (DFT) calculations for the tracer
diffusion~\cite{Gomer} of a single In adatom on a GaAs surfaces. In
particular, we investigate the strain dependence of diffusion in order
to clarify the issues raised above concerning heteroepitaxy of a
strained system. This problem can be viewed as a 2D analog to the
effect of pressure on diffusion in bulk materials.~\cite{Philibert:91}
In a second part, we discuss the impact of these findings for growth
for two typical situations, nucleation on a strained capping layer for
low In concentrations, and diffusion-limited growth of
quasi-one-dimensional islands. While these topics have been discussed
in the literature in the context of thermo-chemical
diffusion,~\cite{Srolovitz,Xie94,Chen96,Xie:95} our focus will be on a
\textit{kinetic} description inspired by the results of our
atomistic calculations.

For a systematic first-principles investigation of the effect of
strain on the diffusivity of an In adatom, we decided to use the
\gaas\ surface~\cite{S-Simkin:89} as a specific example. On top of a
complete As surface layer, the $c(4\times 4)$ reconstruction has rows
of As dimers running in the $[\bar 1 1 0]$ direction, with units of
three As dimers interrupted by a dimer vacancy (see \Fref{fig:1}
below).  We have chosen this reconstruction because it forms the
substrate for the initial stages of InAs deposition for temperatures
$T\lesssim 500^\circ$C (see \Rref{Joyce}).  For very arsenic-rich
growth conditions, In deposition is expected to lead to direct
formation of 3D InAs islands, as has been shown by previous
calculations.\cite{Wang} Thus, we can use this system to study both
the diffusion of the first In atoms on a strained substrate (e.g. a
capping layer with buried islands), as well as diffusion in the
vicinity of an InAs island on the surface.  We note, however, that the
commonly used growth conditions to fabricate quantum dots involve
formation of a InAs wetting layer with reconstructions different from
the $c(4 \times 4)$.  Diffusion of In on this wetting layer will be
addressed in a future publication.

In Sec.~\ref{comput} we outline the underlying computational method. The
mapping of the potential energy surface (PES) for In
diffusion on the unstrained \gaas\ surface is presented in
Sec.~\ref{PES}. In Sec.~\ref{interact} we discuss in great details the
In adatom interaction with the surface As dimers. The effect of strain
is then addressed in Sec.~\ref{strain}.  All microscopic results are
critically examined in Sec.~\ref{results} in order to assess possible
morphological consequences for the growth of strained InAs islands on
GaAs. Finally, a summary and discussion of the results is presented in
Sec.~\ref{sec:last}.

\section{Computational details}
\label{comput}

DFT calculations have proven to be an efficient tool to explore the
elementary processes of crystal growth (see for example
Ruggerone, Ratsch and Scheffler~\cite{Ruggerone}). In the setting employed
here,~\cite{Bockstedte} the substrate is modeled by a slab,
representing the topmost seven atomic layers of the \gaas\ surface,
the bottom layer of which was passivated by pseudo-hydrogen atoms.  A
plane-wave basis set with $E_{\text{cut}} = 10$~Ry energy cutoff was
used in conjunction with \textit{ab initio}
pseudopotentials~\cite{Fuchs:99} and the Perdew-Burke-Ernzerhof
generalized gradient approximation~\cite{PBE} to the exchange and
correlation was employed throughout this study. The integration over
the surface Brillouin zone (SBZ) was performed using a Monckhorst-Pack
set with two special \textbf{k}-points,
$[(\frac{1}{2},0,0),(0,\frac{1}{2},0)]$, equivalent to 64
\textbf{k}-points in the irreducible part of the $1\times1$ SBZ. Thus,
the \textbf{k}-mesh conforms with the one previously used for the
$\beta 2(2\times 4)$ reconstruction.~\cite{Kley}

It is common practice, when addressing adatom diffusion, to map out
the relevant potential energy surface(s) that contains complete
information about the diffusion process. Kley, Ruggerone and
Scheffler~\cite{Kley} have recently argued that the adatom-dimer
interaction on the GaAs(001) surface is a crucial factor for the
proper determination of the PES. In order to set the stage for
introducing strain into the problem, we have scanned the
potential-energy landscape seen by the In adatom, by relaxing the
latter along the surface normal, placing it laterally over a set of
equidistant grid points in the (001) plane and allowing the topmost
six slab layers to freely relax. As a reference for calculating the
binding energy of the adatom we have used the sum of the total energy
of the (properly strained) bare surface and the energy of a free,
spin-polarized In atom.~\cite{note} Geometries were considered
converged when all residual forces were smaller than 0.025~eV/\AA.

The adatom-dimer interaction poses a multidimensional problem, since
not only the adatom itself, but also all degrees of freedom of the
surface atoms are involved in this processes. Therefore, even a full
relaxation starting from an adatom above the surface may only lead to
a local minimum, while other minima may exist that can only be reached
from different starting configurations. Test calculations showed that
the adatom-surface distance and the As-As distance in the dimer are
most important, and the 2D configurational space defined by these
coordinates is suitable to image the In-surface dimer interaction.
Towards this end a special constrained relaxation was carried out
allowing the In adatom and the central As dimer beneath it to be moved
as a rigid unit (see the inset in \Fref{fig:3} below).  The relative
position of these three atoms defines a point in a 2D slice through
the corresponding multidimensional energy hypersurface. Performing the
constrained relaxation, we succeeded in mapping out the 2D PES
governing the In-surface interaction in a point-by-point
fashion. Further details will be given in Sec.~\ref{interact}.

The strain field at a step or island edge varies slowly on the scale
of the lattice constant. Therefore we can treat the inhomogeneously
strained surface by performing DFT calculations with a locally
adjusted lattice parameter.  Accordingly, to investigate the influence
of strain on surface diffusion, the lateral lattice constant $a$ was
uniformly changed in the range of $\pm 8$\% around its value $a_0$
calculated for the unstrained material, thus defining the isotropic
surface strain tensor
$\eps_{\alpha\beta}=\varepsilon\delta_{\alpha\beta},$ with
$\eps=a/a_0-1,$ $\delta_{\alpha\beta}$ being the Kronecker delta,
relaxing again the system as already explained and recording the In
binding energy for the relevant sites on the PES. It is important to
note that any change of the super-cell volume $V$ is accompanied by a
change in the quality of the plane-wave basis set.  To account for
this effect we have corrected the calculated total energy of the
supercell $E_{\rm tot}(E_{\text{cut}},V)$ according to the scaling
hypothesis by Rignanese~\etal.~\cite{Rignanese:95}

\begin{figure}[t]
 \centering
 \mbox{
  \psfig{file=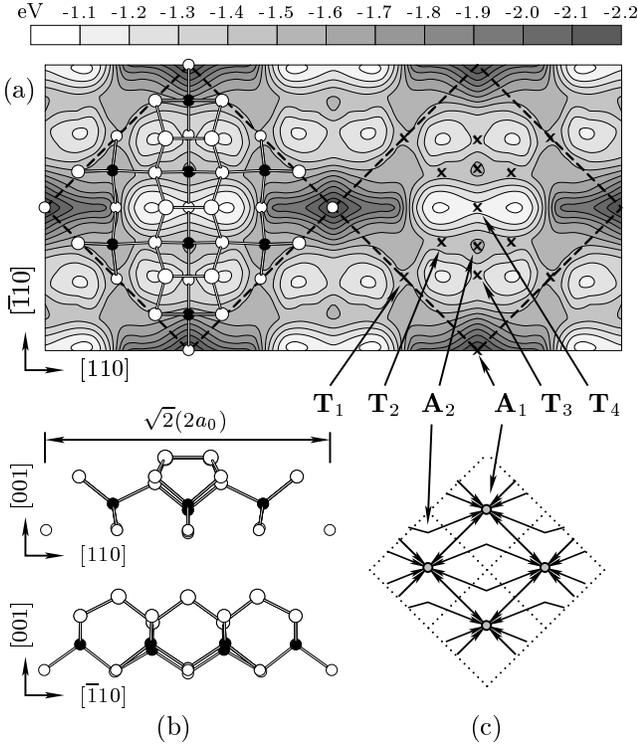,width=\columnwidth}
 }
 \vskip 0.5cm
 \caption{(a) Potential-energy landscape for an In adatom on the
 GaAs(001)-$c(4\times4)$ surface. The adatom is relaxed from 2 \AA\
 above the surface. Contour-line spacing is 0.1 eV; atomic positions in
 the clean surface unit cell are indicated for atoms in the upper four
 layers (As: empty circles; Ga: filled circles), side views shown in
 panel (b), where $a_0$ is the bulk GaAs lattice constant.
 The dashed squares show the surface unit cell.  (c) Sketch of
 the 2D network of sites used in the random walk formalism (four unit
 cells are depicted).
 }
\label{fig:1}
\end{figure}

\section{Indium diffusion on 
 G\lowercase{a}A\lowercase{s}(001)-$\mathbf{\lowercase{\textit{c}}(4 \times 4)}$
}

\subsection{Potential energy surface}
\label{PES}

The mapping procedure resulted in the PES shown in \Fref{fig:1} and
the binding energies of the In adatom at the adsorption sites
(${\mathbf{A}}_i$) and saddle points (${\mathbf{T}}_k$) are given in
\Tref{tab:1}. The In adatom, like Ga/GaAs(001)-$c(4\times4)$
(cf. \Rref{LePage} and \Tref{tab:1}), preferentially adsorbs at the
four-fold coordinated hollow site ${\mathbf{A}}_1$ (the missing dimer
position), where it interacts with the dangling bonds of the second
layer As atoms. Two other very shallow minima ${\mathbf{A}}_2$ are
located in between the center and the two edge dimers. Jumps between
the adsorption sites occur through four symmetry-inequivalent saddle
points ${\mathbf{T}}_k$, with ${\mathbf{T}}_1$ being lowest in energy.

Within transition-state theory~\cite{Hanggi} the hopping rate
between sites ${\mathbf{A}}_i$ and ${\mathbf{A}}_f$, crossing saddle
point ${\mathbf{T}}_k$, is given by the ratio of the partition
functions of the system with the adatom at the equilibrium site, $Z_i
\equiv Z({\mathbf{A}}_i),$ and at the saddle point, $Z_k \equiv
Z({\mathbf{T}}_k),$
\begin{equation}
 \Gamma_{fi} = \frac{1}{2\pi}\frac{\kb T}{\hbar}\frac{Z_k}{Z_i}\; , 
\label{Gamma}
\end{equation}
where $\hbar$ is the Planck's constant.  In the conventional case of
an elastically relaxed surface the Helmholtz free energy $F=-\kb T\ln
Z$ is the proper thermodynamic potential.  \Eref{Gamma} can be cast
into the well known Arrhenius form
\begin{equation}
  \Gamma_{fi} = \Gamma^{(0)}_{fi}
   \exp\{
     -[ E({\mathbf{T}}_k) - E({\mathbf{A}}_i) ] / \kb T
   \},
\label{eq:gamma}
\end{equation}
where $E$ is the (static) total energy of the system, read off the PES
in \Fref{fig:1}. The prefactor $\Gamma^{(0)}_{fi}$ has the form,
\begin{equation}
 \Gamma^{(0)}_{fi} = \frac{1}{2\pi}\frac{\kb T}{\hbar} 
 \exp\left(-\Delta U_{\text{vib}}/\kb T + \Delta S_{\text{vib}}/\kb\right),
\label{eq:gammazero}
\end{equation}
where $\Delta U_{\text{vib}}$ and $\Delta S_{\text{vib}}$ are,
respectively, the associated changes in the vibrational energy and
entropy. $\Gamma^{(0)}_{fi}$ is temperature-independent within the
classical harmonic approximation.

\begin{table}[t]
 \caption{Binding energy $E_b$ (eV) of an In adatom at the sites on
  the \gaas\ surface, denoted in \protect\Fref{fig:1}. For
  comparison, $E_b$ of a Ga adatom at the same sites are
  read off the corresponding potential-energy map calculated by
  LePage~\etal\ in Ref.~\protect\onlinecite{LePage} using the 
  local-density approximation.}
 \begin{tabular}{lcccccc} 
  & \multicolumn{6}{c}{Site} \\ \cline{2-7}
  &${\mathbf{A}}_1$ & ${\mathbf{A}}_2$ & ${\mathbf{T}}_1$ & ${\mathbf{T}}_2$ & 
                            ${\mathbf{T}}_3$ & ${\mathbf{T}}_4$ \\ \hline 
  In & $-2.21$ & $-1.54$ & $-1.56$ & $-1.44$ & $-1.27$ & $-1.17$\\
  Ga & $-3.04$ & $-2.20$ & $-2.54$ & $-2.10$ & $-2.00$ & $-1.90$ 
 \end{tabular}
\label{tab:1}
\end{table}
%
In a simplified picture, the In adatom migrates by a random walk on a
2D square lattice defined by the ${\mathbf{A}}_1$ sites,
\Fref{fig:1}~(c).  However, we account for hops between them via both
${\mathbf{T}}_1$ and
${\mathbf{T}}_2$-${\mathbf{A}}_2$-${\mathbf{T}}_2$ with rates
$\Gamma_{11}$ and $\tilde\Gamma_{11},$ respectively.  Indeed, once the
In adatom has reached the ${\mathbf{A}}_2$ site it needs to overcome a
barrier $E({\mathbf{T}}_2) - E({\mathbf{A}}_2)$ of only 0.1~eV in
order to move towards a neighboring ${\mathbf{A}}_1$ site. As
$E({\mathbf{T}}_2) - E({\mathbf{A}}_2) \lesssim 2\kb T$ for typical
growth temperatures, the adatom is unlikely to equilibrate at the
shallow well ${\mathbf{A}}_2$ before it escapes. Thus it is justified
to use a single rate $\tilde\Gamma_{11}$ for the whole path
${\mathbf{T}}_2$-${\mathbf{A}}_2$-${\mathbf{T}}_2$.

Effective diffusion coefficients can now be extracted by applying the
continuous-time random walk (CTRW) formalism.\cite{Kley,Haus:87} Thus,
it is easily worked out that the diffusion tensor in Cartesian
coordinates ($x\; \|\; [110],$ $y\; \|\; [\bar 110]$) reads
\begin{equation}
 D_{\alpha\beta} = 
 \left(\!\!\begin{array}{cc} 
            D_{[110]} & 0                      \\ 
            0         & D_{[\overline{1}10]}  
           \end{array}\!\right) = 4 a_0^2
 \left(\!\!\begin{array}{cc} 
            \Gamma_{11} + 4\tilde\Gamma_{11} & 0\\
            0         & \Gamma_{11} 
           \end{array}
  \!\!\right).
\label{Dtensor}
\end{equation}
The factor 4 in front of $\tilde\Gamma_{11}$ is partly due to the fact
that there exist two equivalent
${\mathbf{T}}_2$-${\mathbf{A}}_2$-${\mathbf{T}}_2$ paths across the
block of three dimers. Another factor two enters because the In adatom
travels in $[110]$ direction $\sqrt{2}$-times longer distance to reach
a neighboring ${\mathbf{A}}_1$ site than along the path crossing the
saddle ${\mathbf{T}}_1.$ \Eref{Dtensor} thus implies that an isolated
In adatom migrates slightly faster in $[110]$ direction, across the
dimer rows, than along the dimer rows in $[\overline{1}10]$ direction,
with anisotropy ratio $D_{[110]}/D_{[\overline{1}10]} = 1 +
4\tilde\Gamma_{11}/\Gamma_{11}.$ The related diffusion barriers,
entering the rates $\Gamma_{11},$ and $\tilde\Gamma_{11}$ are $\Delta
E=0.65$~eV and $\Delta\tilde E\simeq 0.8$~eV (cf. \Fref{fig:1}~(a),
and \Tref{tab:1}). One gets a rough estimate for the contribution of
the ${\mathbf{A}}_1
\stackrel{{\mathbf{A}}_2}{\longleftrightarrow}{\mathbf{A}}_1$ channel
by assuming that $\Gamma_{11}^{(0)}$ and $\tilde\Gamma_{11}^{(0)}$
differ inessentially. Thus, e.g. at $T=450^{\circ}$C,
$D_{[110]}/D_{[\overline{1}10]}$ would exceed unity by about 50\%.  At
sufficiently low temperatures, however, one should include
${\mathbf{A}}_2$ in the 2D network of sites and consider branching of
the diffusion pathways towards neighboring ${\mathbf{A}}_1$ or
${\mathbf{A}}_2$ sites.  Although an analytic result for
$D_{\alpha\beta}$ can still be derived in this case within the CTRW
formalism, the expressions are rather cumbersome and one has to seek
for simplifications requiring knowledge of all $\Gamma_{fi}$ rates.

In comparison to the results by LePage~\etal~\cite{LePage}
(\Tref{tab:1}) for Ga/\gaas, In appears to diffuse on a less
corrugated PES. Thus, for example, at the ${\mathbf{A}}_1$ site, being
the most stable for both In and Ga, the In adatom is by $\approx 0.8$~eV
less bound than Ga. Furthermore, the $c(4\times 4)$ PES
provides two additional adsorption sites for Ga as compared to In:
between the edge dimers (${\mathbf{T}}_1$ in \Fref{fig:1}) as well as
in between ${\mathbf{A}}_1$ and the center dimer along $[110]$
(cf. Fig.~2 of \Rref{LePage}). At the ${\mathbf{T}}_1$ site, which is
a stable adsorption site for Ga, the latter is by 1.0~eV
more strongly bound than In.  These differences can be easily
rationalized in terms of the differences in the cation-As bond
strength in the corresponding binary compounds (GaAs, InAs) and the
larger ionic radius $R_{\text{\scriptsize{In}}} $ of
indium;~\cite{radii} we note however that part of the differences is
to be attributed to the use of the local density approximation in
\Rref{LePage}. The cohesive energy per cation-As pair is lowest for
InAs ($E_{\text{coh}}^{\text{InAs}} = 6.20$~eV) as compared to GaAs
($E_{\text{coh}}^{\text{GaAs}} = 6.52$~eV) and AlAs
($E_{\text{coh}}^{\text{AlAs}} = 7.56$~eV), see
e.g. \Rref{Bechstedt:88}.  The barriers for diffusion of group-III
cations on the GaAs surface follow the trend given by the binding
energies, as has been also observed in a first-principles study of Ga
and Al diffusion on the GaAs(001)-$\beta2(2\times 4)$
surface.~\cite{Kley}

\subsection{Interaction of Indium with As-As bonds}
\label{interact}

\begin{figure}[t]
 \centering 
 \mbox{
  \psfig{file=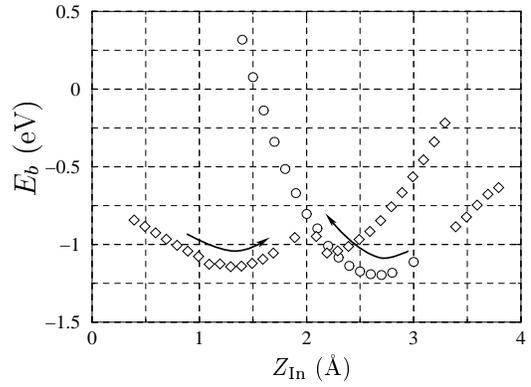,width=0.8\columnwidth}
 } 
 \vskip0.5cm 
 \caption{Binding 
 energy of an In adatom interacting with the center As dimer as a
 function of the $z$-coordinate of the adatom $\zin.$ The latter is 
 measured from the $z$-coordinate of the center dimer for the bare surface.
 Arrows indicate the order in which the calculations were performed:   
 adsorption ($\protect{\circ}$) and desorption ($\protect{\diamond}$).
 }
\label{fig:2}
\end{figure}

Since the $c(4\times 4)$ reconstruction represents a \textit{double}
layer of arsenic, of which the top As atoms form As dimers
(\Fref{fig:1}), the incorporation of In into the cation sublattice
requires the topmost arsenic layer to be eventually replaced by In
atoms.  One obvious way how this incorporation could occur is by
splitting of the As-As bonds in a reaction with an In adatom.  For an
understanding of heteroepitaxy, it is therefore important to study
such processes. Furthermore, Kley, Ruggerone and Scheffler~\cite{Kley}
have pointed out for Ga/GaAs(001)-$\beta2(2\times 4)$ that the adatom
interaction with the surface As dimers has important consequences for
the cation diffusivity, since the broken As-As bond provides a very
stable adsorption site for Ga.  The underlying mechanism has been
identified to be the replacement of the rather weak surface As-As
dimer bond by stronger cation-As bonds, cf.~\Rref{Kley}.  For a valid
description of In diffusion by the PES shown in \Fref{fig:1}, we
therefore have to check if reaction of In with the As-As bonds can
lead to more stable binding sites for In than the minima of the PES.

\begin{figure}[t]
 \centering 
 \mbox{
  \psfig{file=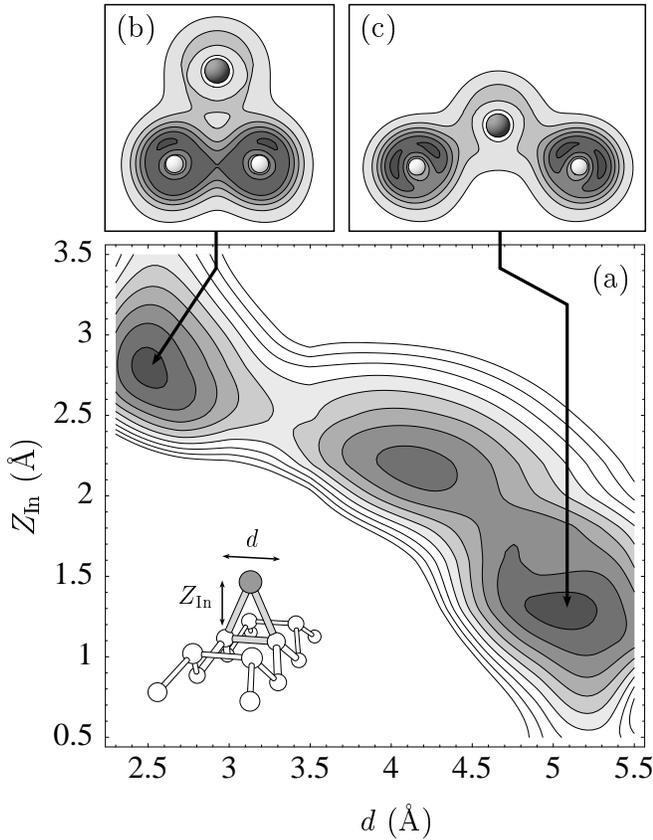,width=\columnwidth}
 } 
 \vskip0.5cm
 \caption{(a) Binding energy of an In adatom interacting with the
 center As dimer as a function of the As-As distance $d$ and the In
 height above the midpoint of the dimer $\zin,$ as indicated in the inset.  
 (b,c)
 Bonding configuration and the valence electron density in the plane
 containing the adatom and the dimer for the two deeper minima of $E_b$.
 }
\label{fig:3}
\end{figure}

We first sample $E_b$ as a function of the adatom height $\zin$ above
the ${\mathbf{T}}_4$ site, see \Fref{fig:2}.  We perform a series of
calculations for various values of $\zin$, where in each calculation
the adatom is kept fixed, while the substrate is allowed to freely
relax.  In subsequent calculations of the series, the geometry of the
substrate atoms from the previous calculation is used as input.  We
find that the outcome of such a series of calculations depends on the
initial geometry.  While a set of data points modeling adsorption,
starting from $\zin=3$~{\AA} above the closed dimer, shows an energy
minimum at $\zin\simeq 2.7$~{\AA}, a series of calculations for
desorption, starting from an adatom incorporated in between the As
dimer atoms at $\zin\simeq 0.5$~\AA, finds a minimum at
$\zin\simeq1.3$~\AA.  Both minima have nearly the same depth,
$E_b\simeq -1.2$~eV.
The corresponding bonding configurations and total valence electron
densities are shown in \Fref{fig:3}~(b) and (c). The discontinuity and
hysteresis in $E_b$ seen in Fig.~\ref{fig:2} due to the dimer opening
or closing indicates that the information gained from the $\zin$
coordinate alone is insufficient for building up a complete picture of
the adatom-dimer interaction.

Therefore we have subsequently imaged $E_b$ in a 2D configurational
space including also the As-As distance $d.$ In these calculations,
both $d$ and the distance between the $z$-coordinate of the adatom and
the As-As midpoint are held fixed while relaxing the remaining
coordinates of the system.  As a result, it becomes evident that the
adatom, approaching the dimer, first goes through a minimum of $E_b$
with the dimer bond being almost intact ($d=2.56$~\AA),
\Fref{fig:3}~(b). Upon further push towards the surface the In adatom
splits the dimer bond, overcoming a barrier of $\simeq 0.35$~eV, and
stays shortly within the $\zin$ channel in a second shallower feature
of the PES. The formation of directed In-As bonds (see
\Fref{fig:3}~(c)) gives rise to a third minimum of $E_b$ at
$(\zin=1.3,\, d=5.1 $~\AA).

\begin{figure}[t]
 \centering 
 \mbox{
  \psfig{file=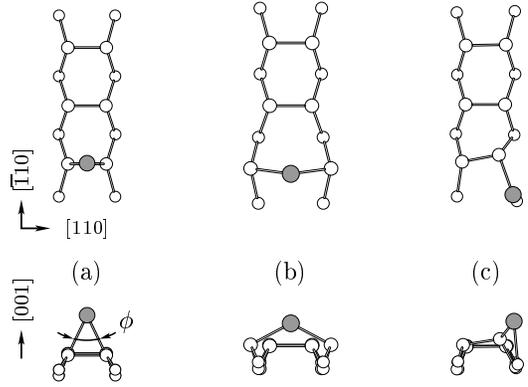,width=0.8\columnwidth}
 } 
 \vskip0.5cm 
 \caption{Different bonding configurations of an In adatom interacting with the
 edge As dimer; atomic positions in the topmost two atomic layers are
 shown (In: shaded circle, As: empty circles). (a) In adatom sitting
 above the closed edge dimer, similar to \protect\Fref{fig:3}~(b)
 (configuration corresponding to the ${\mathbf{T}}_3$ site). $\phi$
 indicates the As-In-As bond angle; (b) In adatom splitting the dimer,
 similar to \protect\Fref{fig:3}~(c); (c) In adatom splitting the
 dimer back-bond.
 }
\label{fig:4}
\end{figure}

In a similar way, we analyze the adatom interaction with the edge
dimers in the $c(4\times 4)$ unit cell, \Fref{fig:4}. Since the edge
dimer has only one neighbor, the second-layer As atoms are expected to
relax more efficiently. Indeed, comparing similar bonding
configurations for the In atom at the center dimer, \Fref{fig:3}~(c),
and at the edge dimer, \Fref{fig:4}~(b), we find 
$E_b = -1.3$~eV 
for the latter, which is only slightly lower than
$E_b({\textbf{T}}_3).$ A third possibility, where In attacks the outer
back-bond of an edge dimer, is shown in \Fref{fig:4}~(c). This
configuration results in $E_b = -1.25$~eV.

It is now clear that additional binding sites for In, related to
broken As-As bonds, are energetically higher than adsorption on the
PES of \Fref{fig:1}, and are therefore not substantially populated in
equilibrium.  Thus, the mechanism operating in the case of
Ga/GaAs(001)-$\beta2(2\times4)$ is absent in the
In/GaAs(001)-$c(4\times4)$ system. Indeed, the more bulky In adatom
with an ionic radius~\cite{radii} $R_{\text{In}}$ larger than that of
Ga, when inserted into the As dimer, introduces substantial elastic
distortion of the dimer As back-bonds that cannot be energetically
compensated by the gain due to rehybridization. Note that even in the
case of an open dimer the In adatom resides 1.3~\AA\ above it,
\Fref{fig:3}~(c), which implies an As-In-As bond angle $\phi \simeq
125^{\circ},$ while the Ga adatom is incorporated almost collinearly
with the two As atoms,~\cite{Kley} $\phi \simeq 175^{\circ}.$ In
summary, our analysis justifies the use of a single PES (\Fref{fig:1})
for In diffusion in the CTRW formalism.

\subsection{Effect of strain}
\label{strain}

The foregoing discussion allows us to single out the main route for
the adatom migration: ${\mathbf{A}}_1
\stackrel{{\mathbf{T}}_1}{\longleftrightarrow} {\mathbf{A}}_1.$ Hence,
the objective in this section is to analyze the influence of surface
elastic strain on the $\Gamma_{11}$ rate.  A non-vanishing strain
field in the substrate results in a different equilibrium
configuration of the topmost atomic layers. Consequently, both the
surface phonon spectrum and the PES will experience changes affecting
in turn both the frequency prefactor and activation energy in the
exponential in \Eref{eq:gamma}.  The net effect of strain is thus
determined by the interplay between the latter two effects.  One may
expect, however, that the dominant contribution comes from variations
in the diffusion barrier $\Delta E \equiv
E_b({\mathbf{T}}_1)-E_b({\mathbf{A}}_1),$ for it enters an
exponential. This motivated us to concentrate mainly on the strain
renormalization of $\Delta E,$ but our approach allows for the
influence of $\Gamma^{(0)}$ to be also incorporated without detailed
knowledge of its functional dependence on strain.

\begin{figure}[t]
 \centering
 \mbox{
  \psfig{file=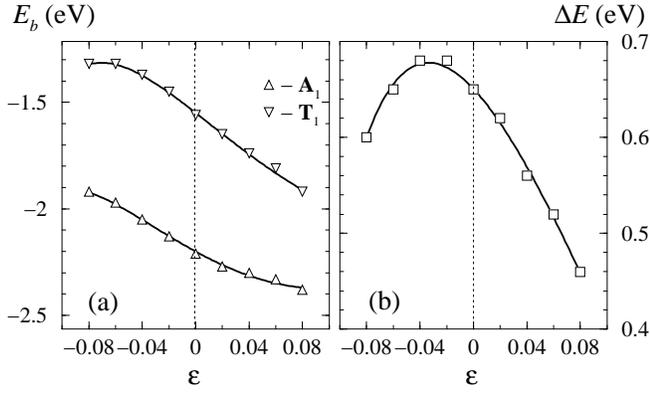,width=\columnwidth}
 }
 \vskip 0.5cm
 \caption{(a) Binding energy $E_b$ as a function of strain $\eps$ for
 an In adatom at the ${\mathbf{A}}_1$ and ${\mathbf{T}}_1$ sites; (b) diffusion
 barrier $\Delta E \equiv E_b({\mathbf{T}}_1)-E_b({\mathbf{A}}_1)$ as a
 function of $\eps$. Full curves on both panels represent least-squares
 polynomial fits to the calculated points.
 }
\label{fig:5}
\end{figure}

For each particular value of $\eps$ the In adatom is placed above the
${\mathbf{A}}_1$ and ${\mathbf{T}}_1$ sites and the same relaxation
scheme as for mapping the PES, Sec.~\ref{PES}, is applied to obtain
the respective binding energy $E_b(\eps).$ The calculated values are
shown in \Fref{fig:5}.  Interestingly, we find that $\Delta E(\eps)$
is a monotonically decreasing function for any tensile strain
($\eps>0$) employed in the calculations, \Fref{fig:5}~(b). To be
specific, this behavior has its onset at $\simeq 3$\% compressive
strain, where $\Delta E$ reaches a maximum of 0.68~eV. Applying larger
compressive strain leads to a reduction of $\Delta E$, with the
$\Delta E(0)$ value recovered again for $\eps= -0.06.$ The
non-monotonic dependence on strain can be rationalized by inspecting
the $E_b(\eps)$ curves, given in \Fref{fig:5}~(a). While for $\eps<0$
$E_b$ at the adsorption site ${\mathbf{A}}_1$ follows a linear law
with a slope of $-3.8$~eV, the binding energy at the saddle point
${\mathbf{T}}_1$ contains, although small, non-linear terms in strain
which do not cancel in the evaluation of $\Delta E.$ For an
inhomogeneously strained sample, the pronounced strain dependence of
$E_b$ for \textit{both} the adsorption site and the saddle point will
introduce a position dependence of $\Delta E.$ This finding complies
with none of the two limiting scenarios of changes of $\Delta E$
discussed in the literature,~\cite{Schroeder:97} where only either
$E_b({\mathbf{A}}_i)$ or $E_b({\mathbf{T}}_k)$ contributes.  We would
also like to emphasize that the commonly employed
linearity~\cite{Dobbs:97b,Koduvely:99} for the strain dependence of
the diffusion barrier $\delta(\Delta E(\eps))$ is not justified in the
case of \inongaas\, as clearly seen from \Fref{fig:5}~(b). Thus one
needs to go to higher order terms in $\eps$ to adequately describe the
observed $\delta(\Delta E(\eps))$ behavior.  It is also important to
point out that strain does not change qualitatively the discussion
about the interaction of the In adatom with As-As bonds, based on the
PES in \Fref{fig:3}. Extensive tests over the entire range of strain
considered here were carried out for $E_b$ of In at the two stable
minima, \Fref{fig:3}~(b) and (c). We found that the binding
configuration of \Fref{fig:3}~(b) was always slightly preferable over
the one in \Fref{fig:3}~(c), but the strongest binding site for In
remains to be ${\mathbf{A}}_1.$

Up to now, the reports in the literature about the effect of strain on
the diffusion barrier are scarce. A slight lowering of the diffusion
barrier upon tensile strain ($\eps>3.5$\%) has been reported for Ag
self-diffusion on a Ag(111) slab within the effective-medium
theory.~\cite{Brune} The first-principles treatment of the same
system~\cite{Ratsch} has found instead a linear increase of the
barrier with strain.  For semiconductors, a lowering of the diffusion
barrier upon tensile strain has only been reported for Si adatom and
dimer diffusion on Si(001),\cite{Roland:92,Spjut:94,Zoethout:00}
although the underlying binding trends inferred from the MD
simulations~\cite{Roland:92} are opposite to those shown in
\Fref{fig:5}~(a).

Given the strain dependence of the diffusion barrier described above,
the basic question arises whether diffusion limitations can be
observed in the growth kinetics of InAs on GaAs.  This would be the
case if the adparticle diffusivity is reduced for relevant material
parameters and growth conditions.

As the substrate around an InAs island, e.g. of pyramidal or truncated
pyramidal shape, is under compressive strain, cf. \Rref{Moll98} and
\Fref{fig:8} below, an indium adatom approaching the island samples
the $\eps < 0$ branch of $\Delta E(\eps)$.  This branch is accurately
described by
\begin{equation} 
 \delta(\Delta E (\eps)) =  \delta E_{\rm max}
 \left[ 1-\left(\frac{\eps}{|\eps_{\rm max}|}+1
 \right)^2\right], \quad \eps < 0.
\label{profile}
\end{equation}
\Eref{profile} gives the excess diffusion barrier over the one for the
unstrained surface $\Delta E(0)$, parameterized by the maximum excess
$\delta E_{\rm max}=30$~meV, and the strain value at which it occurs,
$\eps_{\rm max}=-3$\%.  On the basis of \Eref{profile}, one can write
a rather general expression for the diffusion coefficient taking
account of the effect of strain,
\begin{equation}
 D(\eps) = D_0(1+2\eps)\,\delta\Gamma^{(0)}(\eps) 
 \exp\left[-\frac{\delta(\Delta E (\eps))}{\kb T}\right],
\label{dcoeff}
\end{equation}
where $D_0\equiv\text{const}$ is the value of $D$ for the unstrained
surface, $\delta\Gamma^{(0)}=\Gamma^{(0)}(\eps)/\Gamma^{(0)}(0)$ is
the reduction or enhancement factor of the attempt frequency
$\Gamma^{(0)},$ and $2\eps\equiv{\rm Tr}\,\eps_{\alpha\beta}$ is the
relative change in the surface area.

A first estimate of the expected reduction of $D$ within the typical
temperature range 350--500$^{\circ}$C used for InAs deposition on the
$c(4\times 4)$-reconstructed GaAs(001) substrate~\cite{Joyce} can be
obtained by inserting \Eref{profile} in (\ref{dcoeff}), setting
$\delta\Gamma^{(0)}\equiv 1$.  The resulting reduction,
$D(\eps_{\text{max}})/D_0 \simeq 0.6,$ turns out to be small due to
the smallness of $\delta E_{\text{max}}.$ As a consequence, changes of
the prefactor due to the effect of strain on lattice vibrations are
equally important in determining the strain renormalization of the In
diffusivity on the \gaas\ surface in the relevant temperature regime.

\begin{figure}[t]
 \centering
 \mbox{
  \psfig{file=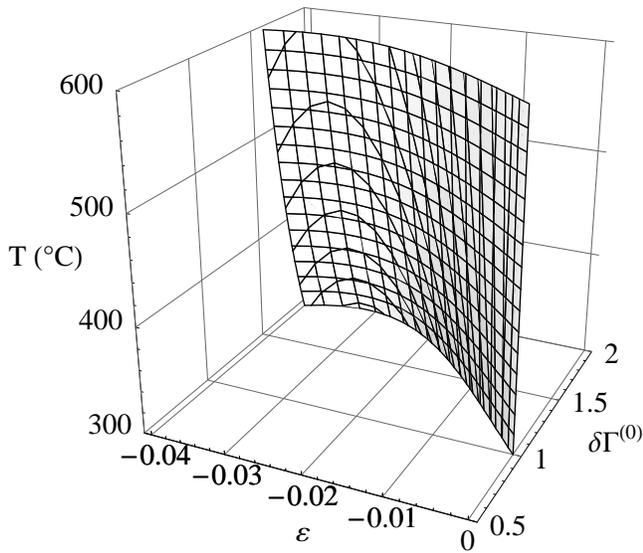,width=\columnwidth}
 }
 \vskip 0.5cm
 \caption{The $D/D_0 = 1$ isosurface in the 3D parameter space
 $(\eps,\delta\Gamma^{(0)},T).$ The view point is from the side where
 $D/D_0<1.$ Points beyond the isosurface correspond to enhanced
 diffusivity $D/D_0>1.$
 }
\label{fig:6}
\end{figure}

Although it is possible to obtain $\delta\Gamma^{(0)}$ from
first-principles calculations, it is difficult to get an estimate that
is better than a factor of two with reasonable computational effort.
Hence we consider $\delta\Gamma^{(0)}$ as an independent parameter in
the following analysis.  Thus the right-hand side of \Eref{dcoeff} is
a function of three parameters, $\eps,\delta\Gamma^{(0)},$ and $T.$
The region in parameter space where strain-induced growth limitations
can be expected is defined by the requirement
\begin{equation}
  D(\eps;\,\delta\Gamma^{(0)}, T)/D_0<1.
\label{domain}
\end{equation}
\Fref{fig:6} represents the isosurface $D/D_0=1$ in the 3D parameter
space of $\eps,$ $\delta\Gamma^{(0)},$ and $T.$ We note that 
the reduction in diffusivity due to positive
$\delta(\Delta E),$ especially for lower temperatures, persists even for
$\delta\Gamma^{(0)} > 1,$ i.e. in presence of the so-called
compensation effect.~\cite{Boisvert:95} However, for a very strong 
compensation effect, $\delta\Gamma^{(0)}(\eps)
>\delta\Gamma^{(0)}_{\text{c}} \simeq 2$, no decrease in $D$  
in the relevant range of $\eps$ and $T$ values can be expected.

Finally, we have performed DFT calculations to obtain an estimate of
$\Gamma_{11}^{(0)}(-0.04)$, using the harmonic approximation for the
lattice vibrations and a force-constant matrix involving only the
degrees of freedom~\cite{Ratsch:98} of the In adatom. This estimate
indicates $\sim 70$\% increase of the prefactor for 4\% compressive
strain.  In this case one cannot expect more than a few percent
maximum reduction of $D$, which would in turn make diffusion
limitations for the specific example of the \gaas\ surface rather
unlikely.

\section{Consequences for growth}
\label{results}

It is interesting to discuss the impact of strain on the growth
kinetics of both 2D and 3D arrays of self-assembled strained islands
in a more \emph{general} context.  As already mentioned in the introduction,
for these two situations different regimes of strain are realized, and
we address them separately.

\subsection{Growth on a capping layer with buried islands}

In analogy to the previous discussion one can also consider diffusion
in the regime of tensile strain, $\eps > 0.$ This situation is
pertinent to the heteroepitaxial growth of 3D arrays of vertically
self-organized QDs. When a 2D sheet of QDs is completed and capped by
a spacer GaAs layer, the GaAs lattice is expanded in the regions above
the buried InAs QDs.  We consider the onset of In deposition onto the
spacer layer before nucleation of the first new islands.  A stationary
concentration of adatoms on the surface builds up by the equilibrium
between supply from an atomic In beam source and loss due to
evaporation of In. However, the concentration may vary laterally along
the surface.  In the stationary state the local concentration
$n({\mathbf{r}}_{\|})$ is given by local equilibrium,
$$
n({\mathbf{r}}_{\|})=n_0\exp[-U({\mathbf{r}}_{\|})/\kb T],
$$ 
where $U({\mathbf{r}_{\|}})$ is the binding energy of the In adatoms
at site ${\mathbf{A}}_1,$ and ${\mathbf{r}_{\|}}$ is the coordinate
within the surface.  $U({\mathbf{r}_{\|}})$ is a function of local
strain, as given by $E_b(\eps(\mathbf{r}_{\|}))$ in
\Fref{fig:5}~(a). As can be seen from this figure the binding strength
increases with increasing strain. Thus, the local concentration of
adatoms, and hence the nucleation probability for a new island, is
increased in the region above a buried island where the capping layer
surface is under tensile strain. Our calculations, thus, provide a
microscopic foundation for the frequently made
assumption~\cite{Tersoff:96,Holy:99} that it is easier to nucleate an
InAs island on those regions of the capping surface where the GaAs
lattice constant is widened up and thus more closely matches the InAs
lattice constant.

\subsection{Diffusion limitations in island growth kinetics}
\label{1Dexample}

The conditions under which kinetic growth limitations can be expected
were discussed already in Sec.~\ref{strain} (see \Eref{domain} and
\Fref{fig:6}).  It is interesting to illustrate the possible
consequences of such limitations for the island sizes. This is pursued
here within the framework of a simple model problem based on the flat
island approximation.~\cite{Tersoff:93} One might think, for instance,
of adatom diffusion towards the extended edge of a quantum wire.

As a first step we address the strain renormalization of the
adparticle diffusivity due to an isolated island. The strain field it
creates in the underlying substrate surface, within the adopted
model, has the form
\begin{equation}
  \eps (x) = \eta\,\ln\!\left| \frac{P_2(x)}{Q_2(x)}\right|.
  \label{1dstrain}
\end{equation}
The island geometry (height $h,$ width $s,$ and the tilt angle
of the side facets $\theta$) completely determines the coefficients in
the second-order polynomials $P_2(x)$ and $Q_2(x),$ whereas the
elastic properties of the material system (e.g. Poisson ratio, shear
modulus of the substrate) enter the prefactor $\eta.$

\begin{figure}
 \centering
 \mbox{
  \psfig{file=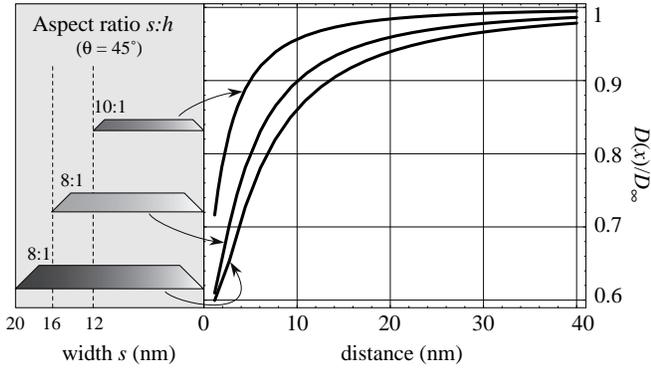,width=\columnwidth}
 }
 \vskip0.4cm
 \caption{Strain renormalization of $D(x),$ according
 to Eqs.~(\protect\ref{dcoeff}) and (\protect\ref{1dstrain}), versus island
 size within the 1D flat island approximation at $T=450^{\circ}$C and
 absence of compensation effect, $\delta\Gamma^{(0)} =1$. $D_{\infty}$
 refers to the asymptotic value of $D$ at infinitely large distance
 from the island. The tilt angle of the island side facets  is
 $\theta$ and $h$ is their height.  The island/substrate system is
 assumed to be InAs/GaAs.
 }
\label{fig:7}
\end{figure}                                                                   

To assess quantitatively the role of diffusion limitations, we insert
the numerical values for $\delta E_{\rm max}$ and $\eps_{\rm max}$
obtained in Sec.~\ref{strain}.  For three islands of different size
\Fref{fig:7} shows the spatial dependence of $D(x)$ obtained by
inserting \Eref{1dstrain} into \Eref{dcoeff}, for $\delta\Gamma^{(0)}
=1$ and $T=450^{\circ}$C. As seen from the short-range behavior of the
diffusion coefficient, the larger islands can be about 20--30\% more
effective in hindering the adatom migration provided that \textit{no
substantial compensation effect} from the $\Gamma^{(0)}$ prefactor is
present.  Consequently, as long as the islands grow via
strain-dominated surface mass transport and the last mentioned
condition is met, the compressive strain field may lead to retarded
growth of the larger islands.

\begin{figure}
 \centering
 \mbox{
  \psfig{file=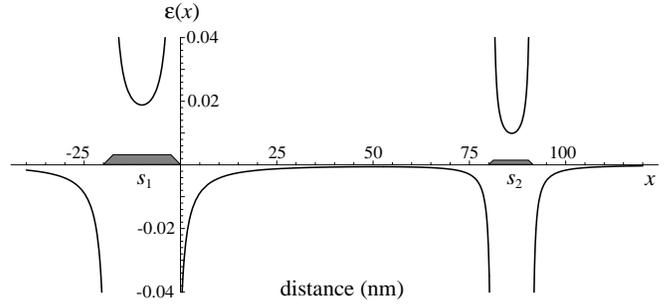,width=\columnwidth}
 }
 \vskip0.4cm
 \caption{Strain field $\eps(x)$ at the substrate surface, according
 to~\protect{\Eref{1dstrain}}, for two 1D islands of width $s_1$ and
 $s_2$ whose edges are separated by distance $L$. The island/substrate
 system is InAs/GaAs.
 }
\label{fig:8}
\end{figure}   

Consider now two islands of characteristic size $s_1$ and $s_2,$ with
$s_1 > s_2,$ separated by a distance $L\gg
s_1,s_2,$~\Fref{fig:8}. Supply of adparticles to the surface is
maintained by a stationary flux $F_0.$ One may ask then, what is the
steady-state adparticle density distribution $n(x)$ at the surface,
and how does it affect the diffusional currents of single adatoms
toward the islands, $-j_1$ and $j_2$?  This is a standard problem in
kinetics;~\cite{BCF,Tersoff:97} however, we require it to be solved
for a spatially varying migration potential $U(x)$ due to the presence
of strained islands. Again, we can exploit the results obtained in
Sec.~\ref{strain}, identifying $U(x)$ with $E_b(\eps(x))$ for the
adsorption site ${\mathbf{A}}_1$.  From \Fref{fig:5}~(a) it becomes
clear that a compressive strain of a few percent significantly weakens
the binding of the In adatom at the ${\mathbf{A}}_1$ site. Thus the
coherently strained island gives rise to a repulsive potential $U(x)$
that amounts to a few tenths of an eV. The time-dependent single atom
density $n(x,t)$ satisfies a Smoluchowski-type
equation~\cite{Festa:78,Risken:96} that takes explicit account of the
field of force due to $U(x),$
\begin{equation}
  \frac{\partial n}{\partial t} =
  \frac{\partial}{\partial x}\left[  
    D(x)\left(\frac{\partial n}{\partial x} +\frac{n}{\kb
  T}\frac{dU(x)}{d x} \right)
  \right] + F_0
\label{smol}
\end{equation}

In the simplest case, when the island edges act as perfect sinks,
i.e. $n(0)=n(L)=0,$ the stationary solution reads as
\begin{equation}
 n(x)=F_0 e^{-\frac{U(x)}{\kb T}}\!\!\int\limits_0^x\frac{x_0-x'}{D(x')}
      e^{U(x')\over k_{_{\rm B}}T} dx',
\label{solution}
\end{equation}
with $0<x_0<L$ being the position between the two islands where the
total adparticle current vanishes $j(x_0)\propto -\nabla n|_{x_0}=0,$
\begin{equation}
 x_0 = \left[\int\limits_0^L\frac{1}{\tilde D(x)}dx\right]^{-1}
       \int\limits_0^L \frac{x}{\tilde D(x)}dx,
\label{x0}
\end{equation}
with $\tilde D(x) = D(x)\exp[-U(x)/k_{_{\rm B}}T].$ It is now
straightforward to obtain the result that relates $j_1$ with $j_2$
\begin{equation}
 \left| \frac{j_1}{j_2}\right| = \frac{x_0}{L-x_0}.
\label{ratio}
\end{equation}

Without the effect of strain, the adatom density has a simple
parabolic profile,
\begin{equation}
 n_0(x) = \frac{F_0}{2D_0} (2x_0 - x) x
\label{n0}
\end{equation}
with its maximum being exactly at the midpoint between the two
islands, $x_0=L/2.$ The strain renormalization of diffusion shifts
$x_0$ towards the bigger island, thus reducing the particle current
reaching this island.  This simple 1D model problem demonstrates that
the smaller island will grow faster, until $x_0$ gets shifted back
towards the midpoint when the sizes of the two islands have become
equal.  As a consequence, the strain-limited adatom diffusion will
tend to equalize the island sizes by controlling the capture areas for
the two islands competing for the deposited material.  We note,
however, that this effect will be reduced if a compensation from the
frequency prefactor $\Gamma^{(0)}$ is operative.

\section{Summary and discussion}
\label{sec:last}

We have presented the first \textit{ab initio} analysis of the effect
of strain on adatom diffusivity in the context of the heteroepitaxial
growth of InAs QDs on GaAs(001).  In particular, we quantified the
strain dependence of the diffusion barrier for an indium adatom on the
\gaas\ surface. The In interaction with the surface As dimers was also
given due account.  A simple 1D model problem was employed to
demonstrate that the strain-limited diffusion contains an archetype of
self-limiting growth of strained islands. We note that the
self-limiting effect is an intrinsic feature for a system with
spatially varying diffusion coefficients, as indicated by
Eqs.~(\ref{x0}) and (\ref{ratio}), bearing
reduction within a certain operative range of strain values.  For the
lattice-mismatched heteroepitaxy of semiconductor nanostructures, the
strain effect will give rise to a significant repulsive interaction
between a strained island and an adatom diffusing towards the
island. Moreover, the diffusivity will be reduced as well, but a very
accurate treatment is needed to assess its reduction. Our atomistic
calculations yielded a maximum increase of the In diffusion barrier on
\gaas\ of 30~meV for compressive misfit strain.  Since this value is
of the order of $\kb T,$ conclusions about the relevance of this
effect would require as well an accurate calculation of the prefactor
$\Gamma^{(0)},$ which is beyond our present goal.

Finally, we would like to comment on other possible extensions of the
work presented here.  The discussion presented in this article refers
to diffusivity of a single adatom, the so-called \textit{tracer}
diffusion coefficient~\cite{Gomer} (sometimes denoted in the
literature as $D^*$).  Our microscopic results could, however,
serve as an input, for example, to Monte Carlo simulations in order to
shed more light on the effect of strain on adatom diffusivity.
Experimentally, the self-assembled coherent islands are usually grown
on a wetting layer, and thus the In adatoms, in fact, diffuse on an
InAs wetting layer or on one of mixed (In,Ga)As composition. Therefore
the \inongaas\ system used here is suitable for modeling the arrival
of the first In atom to the GaAs surface in the initial stages of InAs
deposition, but the results cannot be taken over directly to the later
stages of island growth. Future research will therefore have to
consider diffusion on a wetting layer and possible account of
alloying.\cite{future}

\acknowledgments

One of the authors (E.~P.) is much indebted to Prof. T.~Mishonov for
the fruitful discussions, his suggestions and help during this study.
We thank also Prof.~E. Sch\"oll, S.~Bose and M.~Meixner for extensive
discussions. This work was partially supported by Deutsche
Forschungsgemeinschaft (Sfb~296).


\end{document}